\documentclass[fleqn,usenatbib]{mnras}

\usepackage{newtxtext,newtxmath}
\usepackage[T1]{fontenc}

\DeclareRobustCommand{\VAN}[3]{#2}
\let\VANthebibliography\thebibliography
\def\thebibliography{\DeclareRobustCommand{\VAN}[3]{##3}\VANthebibliography}

\usepackage{graphicx}	% Including figure files
\usepackage{amsmath}	% Advanced maths commands
\usepackage{bm}
\usepackage{soul}
\usepackage{changes}
\usepackage{cancel}
\newcommand{\sgra}{\mbox{SGR J1550$-$5418~}}
\newcommand{\sgranosp}{\mbox{SGR J1550$-$5418}}

\newcommand{\sgrb}{\mbox{SGR J1935$+$2154~}}
\newcommand{\sgrbnosp}{\mbox{SGR J1935$+$2154}}

\title[Magnetar burst dynamical stability]{%Quantifying randomness: similarities in time-energy behavior of magnetar flares and fast radio bursts
Quantifying chaos and randomness in magnetar bursts
}

\author[Yamasaki, G{\"o}{\u{g}}{\"u}{\textcommabelow s} \& Hashimoto]
{Shotaro Yamasaki\thanks{E-mail: shotaro.s.yamasaki@gmail.com}$^{1}$, Ersin G{\"o}{\u{g}}{\"u}{\textcommabelow s}$^{2}$, and Tetsuya Hashimoto$^{1}$
\\
$^{1}$Department of Physics, National Chung Hsing University, 145 Xingda Rd., South Dist., Taichung 40227, Taiwan\\
$^{2}$Sabanc{\i} University, Faculty of Engineering and Natural Sciences, Tuzla 34956, {\.{I}}stanbul, Turkey
}
\date{Accepted XXX. Received YYY; in original form ZZZ}

\pubyear{2023}

\begin{document}
\label{firstpage}
\pagerange{\pageref{firstpage}--\pageref{lastpage}}
\maketitle

\begin{abstract}
In this study, we explore the dynamical stability of magnetar bursts within the context of the chaos-randomness phase space for the first time, aiming to uncover unique behaviors compared to various astrophysical transients, including fast radio bursts (FRBs).  We analyze burst energy time series data from active magnetar sources \sgra 
and \sgrbnosp, focusing on burst arrival time and energy differences between consecutive events. 
We find a distinct separation in the time domain, where magnetar bursts exhibit significantly lower randomness compared to FRBs, solar flares, and earthquakes, with a slightly higher degree of chaos. In the energy domain, magnetar bursts exhibit a broad consistency with other phenomena, primarily due to the wide distribution of chaos-randomness observed across different bursts and sources. %both magnetars exhibit random behavior that falls between the characteristics of earthquakes and solar flares, implying a complex interplay of crustal and magnetospheric processes during burst triggers. 
%In terms of chaotic behavior, bursts from \sgrb display patterns reminiscent of solar flares, while \sgra events exhibit a dual behavior resembling both solar flares and earthquakes, suggesting varied chaotic characteristics across different episodes and/or magnetar sources. 
Intriguingly, contrary to expectations from the FRB-magnetar connection, the arrival time patterns of magnetar bursts in our analysis do not exhibit significant proximity to repeating FRBs in the chaos-randomness plane.  This finding may challenge the hypothesis that FRBs are associated with typical
magnetar bursts but indirectly supports the evidence that FRBs may primarily be linked to special magnetar bursts like peculiar X-ray bursts from \sgrb observed simultaneously with Galactic FRB 200428.
\end{abstract}

% Don't make up new ones.s

%%%%%%%%%%%%%%%%%%%%%%%%%%%%%%%%%%%%%%%%%%%%%%%%%%

%%%%%%%%%%%%%%%%% BODY OF PAPER %%%%%%%%%%%%%%%%%%

\section{Introduction} \label{sec:intro}
Neutron stars with exceptionally powerful magnetic fields, known as magnetars \citep{DT1992,Paczynski1992}, exhibit magnetic strengths exceeding the Schwinger limit at $B_{\text{cr}}\equiv m^2_e c^3/(\hbar e) \approx 4.4 \times 10^{13}$ G. They have rotation periods spanning seconds and occasionally release recurrent, brief yet tremendously energetic X-ray bursts \citep{Kaspi2017,Enoto2019}.  
Various models have been proposed to explain magnetar burst triggers, some emphasizing internal mechanisms like MHD instabilities within the core or the fracturing of the rigid stellar crust, resulting in the sudden release of magnetic energy from the stellar interior into the magnetosphere \citep{TD1995, TD1996, TD2001}. Others propose external mechanisms involving magnetic reconnections \citep{Lyutikov2003, Gill2010, Yu2012, Parfrey2013}. However, it remains still unclear how these bursts are triggered. 

Intriguingly, magnetar bursts might have a deeper connection with another cosmic puzzle -- the cosmological fast radio bursts (FRBs) \citep{Popov2010,Lyubarsky2014,Pen2015,Cordes2016,Katz2016,Murase2016,Kashiyama2017,Beloborodov2017,Metzger2017,Kumar2017,Wadiasingh2019}, enigmatic extragalactic transient events known for their intense and brief flashes of radio emissions \citep{lorimer07,petroff22}. A potential link between FRBs and magnetars gained more support with the detection of an FRB-like event from the Galactic magnetar \sgrb \citep{Chime,Stare2}, occurring simultaneously with magnetar bursts \citep{integral_paper,hxmt_paper,konus_paper,agile_paper}.
Nevertheless, the trigger mechanism and radiation process behind these phenomena remain subjects of intense debate \citep{zhang20,Lyubarsky2021}.

%\citep{SangLin22}

%\citep{SangLin23}

%\citep{WeiZhaoWang22}

%\citep{Shen23}

%\citep{Katz23}

%\citep{Aschwanden22}

%\citep{WeiWuDai21}

%\citep{Totani2023}

%Numerous bursts, exceeding several thousands in total, have been observed from a handful of repeating FRB sources.
%Analyzing the temporal patterns and energy distributions of these bursts could yield valuable insights into the underlying burst production mechanisms...

Extensive research has explored the statistical properties of magnetar bursts, particularly focusing on the examination of the burst energy distribution \citep{Cheng1996,Gogus99,Gogus00,Woods2006,Nakagawa2007,Collazzi15,Lin2020}. %\sy{\citet{Wang2017} found that the frequency distributions of peak flux, fluence, duration, and waiting time for FRB 121102 are similar to magnetar bursts for the first time. 
Recent studies found that the statistical characteristics of the repeating FRB 121102 (such as the frequency distributions of peak flux, fluence, duration, and waiting times) align closely with those of magnetar bursts \citep{Wang2017,Cheng2020}.

\begin{figure*}
    \centering \includegraphics[scale=0.65]{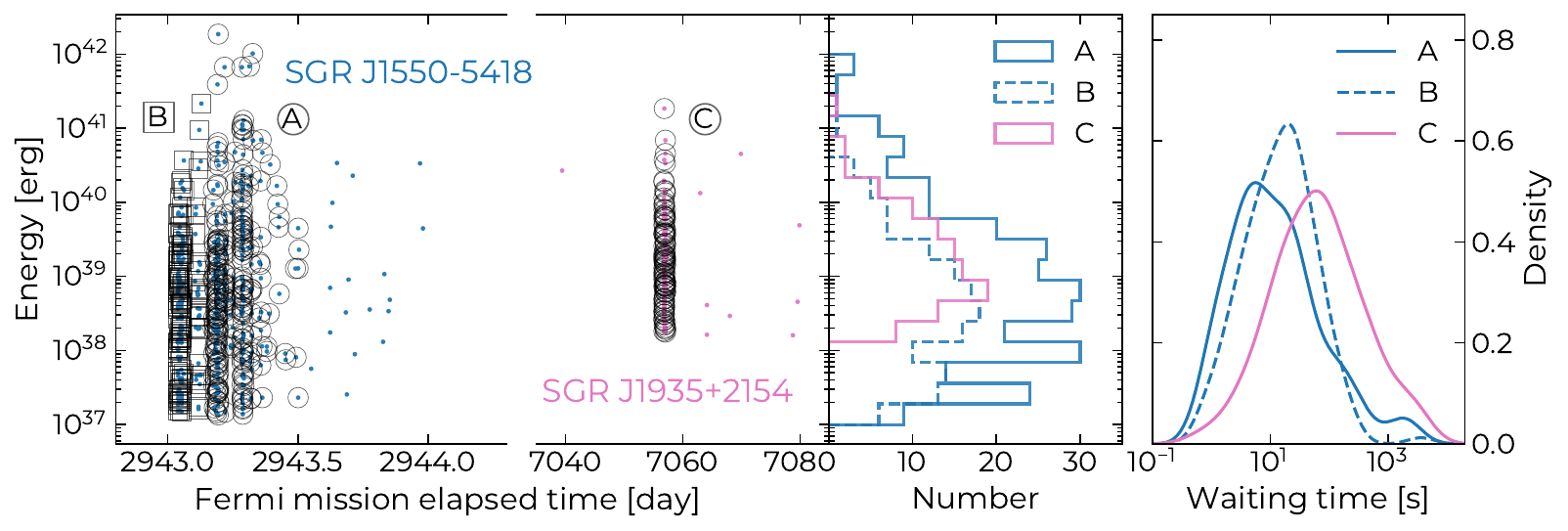}
    \caption{Magnetar bursts from sources \sgra (blue) and \sgrb (pink).  {\it Left panel}: burst energy as a function of burst arrival times (in units of {\it Fermi} mission elapsed time) and energy distributions. The burst groups selected for the dynamical analysis are denoted by black square and circle markers ; {\it right panel}: waiting time distributions.}
    \label{fig:time_series}
\end{figure*}

Recently \citet{Zhang2023} (hereafter Z23) introduced the concept of the Pincus Index and Lyapunov Exponent into the analysis of dynamical stability in astrophysical phenomena. This framework could serve as a new window into the interplay of chaos and randomness, offering a canvas to explore various astrophysical events. 
They compared repeating FRB bursts with diverse physical phenomena such as pulsars, earthquakes, solar flares, and Brownian motion. Their conclusion suggested that FRBs share similarities with Brownian motion in the randomness-chaos phase space. 
However, a crucial piece of the puzzle remained uncharted — the behavior of magnetar bursts within this phase space. Given the potential significance of magnetar bursts as FRB progenitors, understanding their dynamics becomes important. 
Our study, therefore, extends the scope of the investigation to explore the randomness-chaos characteristics of magnetar bursts, %which are a prominent candidate for the origin of FRBs. 
a first step towards unraveling their role as potential progenitors of FRBs. The strength of this new approach lies in its ability to analyze relatively rare transient events such as magnetar bursts. This capability enables the study of sets of magnetar bursts even with relatively limited statistics.
%This investigation not only enriches our comprehension but also aligns with the research pursuits of the FRB community. %The power of this approach lies in its ability to examine infrequent transient events. This enables us to study two particular magnetars, each exhibiting an order of magnitude of around 100 bursts.

%In this letter, we conduct a comparative investigation into the dynamical stability of magnetar bursts and fast radio bursts.  
This letter is structured as follows: in \S \ref{sec:data}, we outline the magnetar burst data and observations utilized in this study. In \S \ref{sec:method}, we detail the analysis of dynamical stability for magnetar bursts. Our findings are discussed in \S \ref{sec:discussion}, followed by a summary of the work.

\section{Data} \label{sec:data}

We used burst data collected with Gamma-ray Burst Monitor \citep[GBM, see][]{Meegan09} on board {\it Fermi} Space Gamma-Ray Telescope (in short, Fermi) from two prolific magnetars: \sgra and \sgrbnosp. The spectra of typical magnetar bursts cut off rapidly above 100 keV. For this reason, we only employed data collected with sodium iodide (NaI) scintillators which are sensitive to photon energies in the range from $\sim$8 keV to about an MeV.

\sgra was identified as a magnetar based on its spin and spin-down rate measurements in radio waveband \citep{Camilo07}. Intriguingly, recent research has suggested that it might have emitted FRB-like bursts in association with an X-ray burst \citep{Israel2021}.
Its first X-ray bursting episode was observed in 2008 October \citep{vonKienlin08} that lasted only a few days. On 2009 January 22, this source entered into its most active bursting episode to date, emitting hundreds of short bursts \citep{vanderHorst12}, as well as a few longer duration and more energetic events \citep[see e.g.,][]{Mereghetti09}. \cite{vanderHorst12} performed a search for untriggered bursts, besides those already triggered GBM detectors. They found 555 bursts (both triggered and untriggered) only on 2009 January 22 whose energy fluences ranged from about 6$\times$10$^{-8}$ erg cm$^{-2}$ to slightly above 10$^{-5}$ erg cm$^{-2}$.  %\dsy{closely examined these events, determined the pairs of bursts with multiple episodes, and} 
Out of 555 bursts identified, the time separations between the onsets of 112 pairs of bursts were shorter than 0.5 s (that is, a quarter of the spin period of \sgranosp). Therefore, they were considered as the pairs of peaks of multi-episodic bursts. As a result. we 
identified 443 bursts for our investigations here. We assumed the distance of 5 kpc \citep{Tiengo10} to determine the isotropic energy release of these events.

\sgrb was discovered in 2014 by exhibiting energetic X-ray bursts. Its spin period and period derivative measurements in X-rays yielded an inferred dipole magnetic field strength of 2.2$\times$10$^{14}$ G, therefore, establishing it a magnetar \citep{2Israel1935}. The source went into burst active episodes again in 2015, 2016 \citep{Lin2020} and 2019 \citep{LinLin2020-2}. On 2020 April 27, \sgrb entered into its most active bursting episode, emitting hundreds of bursts \citep[see e.g.,][]{LinLin2020-2,Younes2020}, including a burst storm \citep{Kaneko2021}. Only a few hours after the onset of this activation, \sgrb emitted the first Galactic FRB \citep{Chime,Stare2} associated with an energetic X-ray burst \citep{integral_paper,hxmt_paper,konus_paper,agile_paper}. For our investigations in this study, we selected 141 bursts that occurred during its 2019 and 2020 activity episodes due to the fact that there were a large number of densely clustered events with fluences in the range from $1.7\times10^{-8}$ to $1.9\times10^{-5}$ erg cm$^{-2}$. We assumed a distance of 9 kpc \citep{Zhong2020} to obtain the isotropic burst energies.

These bursts occurred during multiple observations, resulting in substantial information gaps between them. Therefore, it is crucial that a pair of successive bursts occur within the same Fermi observing (orbital) window. Considering Fermi's low Earth orbit with a period of approximately 96 minutes, the waiting times between successive bursts in the same observing window should not exceed approximately 50 minutes. Based on this criterion, we categorize all bursts into sets of time series corresponding to each uninterrupted observing session. To maximize the statistical yield, we carefully select burst groups with the highest total event counts for each magnetar source. We have identified two data sets labeled as A ($280$ bursts) and B ($145$ bursts) for \sgra and a single data set labeled as C ($105$ bursts) for \sgrb.
Figure \ref{fig:time_series} summarizes the burst arrival time and energy for these two sources, as well as their distributions. The waiting times peak at around $10$--$100$ s and the energy range spans from $10^{38}$ to $10^{42}$ erg,  which is typical for the predominant class of magnetar bursts, often referred to as short bursts.

\section{Dynamical Stability Analysis
} \label{sec:method}
%\subsection{Energy Clustering} 

For each magnetar's energy time series data set described in \S \ref{sec:data} (A, B and C), we investigate a sequence of time differences (or waiting times) and energy differences between two consecutive events, denoted as $\Delta T_i = T_{i+1}-T_{i}$ and $\Delta E_i= E_{i+1}-E_{i}$, respectively. We focus on $\Delta E$, rather than just $E$, because the energy fluctuations within a sequence of bursts could play a crucial role in determining the stability and the transitions between different states within various dynamic systems.
Our dynamical stability analysis in this section covers both time, $\Delta T_i$, and energy, $\Delta E_i$, spaces. 
While our methodology is primarily based on the approach presented in Z23, we compute the relevant quantities (detailed in \S \ref{subsec:PI} and \S \ref{subsec:LLE}) separately for both time and energy spaces within each data set. This is in contrast to the averaging approach employed in Z23, and we provide a reason for this with a demonstration in Appendix \ref{sec:caution}.

\subsection{Pincus Index} 
\label{subsec:PI}

Here, we define Approximate Entropy (ApEn), which is a statistical measure that assesses the degree of randomness within a data series by counting patterns and their repetitions \citep{Pincus1991}. Consider a time series $\mathbf{u}=\{u_1,u_2,\ldots,u_N\}$ with length $N$. In the context of ApEn analysis, we introduce the following parameters:
\begin{itemize}
    \item $m$: a positive integer that represents %the length of a run of data 
    the length of the compared patterns in data with $m\leq N$.
    \item $r$: a positive real number specifying the tolerance or effective noise filter.
    \item $n$: defined as $n=N-m+1$.
\end{itemize}
For each $i$ where $1\leq i\leq n$, we define $\mathbf{x}_i$ as a vector of length $m$: $\mathbf{x}_i=\{u_i,u_{i+1}
,\ldots,u_{i+m-1}\}$.
In other words, $\mathbf{x}_i$ encapsulates a consecutive run of data starting with $u_i$ and comprising $m$ elements.
With these definitions, ApEn for a sequence $\mathbf{u}$ is defined as follows:
\begin{eqnarray}
    {\rm ApEn}\,(m,r;\,\mathbf{u})=\phi^m(r)-\phi^{m+1}(r),
\end{eqnarray}
where
\begin{eqnarray}
    \phi^m(r)&=&\frac{1}{n}\sum_{i=1}^{n}\ln C_i^m(r)\ ,\\
    C_i^m(r) &=& \frac{1}{n}\sum_{j=1}^{n} \theta(d(\mathbf{x}_i^m,\mathbf{x}_j^m)-r)\ ,
\end{eqnarray}
where the Chebyshev distance, denoted as $d$, between $\mathbf{x}_i$ and $\mathbf{x}_j$ is determined by the largest absolute difference between corresponding elements across the vectors, and $\theta$ represents the step function.
Said differently, ApEn quantifies the likelihood of maintaining proximity between pairs of points ($\mathbf{x}_i, \mathbf{x}_j$) in an $m$-dimensional space, given that they are within a distance $r$ of each other. Low ApEn values suggest the presence of patterns, implying some level of predictability in the series, while high ApEn values indicate randomness and unpredictability.

Varying the choice of $m$ significantly influences the computed ApEn values. To account for this, our methodology explored various distance threshold values ($r$) and selected the highest ApEn value, referred to as the Maximum Approximate Entropy (ApEn$_{\rm max}$). This approach effectively mitigates the potential impact of varying $m$ selections on ApEn$_{\rm max}$ outcomes.
Nevertheless, relying solely on ApEn$_{\rm max}$ for cross-comparison across diverse phenomena has limitations. To address this, Z23 introduced the Pincus Index (PI; \citealt{DB2019}), designed to gauge randomness by evaluating the discrepancy in ApEn$_{\rm max}$ prior to and after shuffling sequence elements in $\mathbf{u}$ as follows:
\begin{eqnarray}
    {\cal PI}=\frac{{\rm ApEn}_{\rm max}(m;\mathbf{u}_{\rm original})}{{\rm ApEn}_{\rm max}(m;\mathbf{u}_{\rm shuffled})},
\end{eqnarray}
This normalization method allows for comparisons across different phenomena. In this analysis, we maintained $m=2$ to ensure consistency with Z23. 
To compute PI, we conducted $10^3$ shuffling iterations of $\mathbf{u}_{\rm original}$ and calculated the mean of each ${\rm ApEn}_{\rm max}(m;\mathbf{u}_{\rm shuffled})$. The associated error was calculated based on the standard deviation of the PI distribution. We computed two PI values for $\Delta T_i$ and $\Delta E_i$ separately, resulting in the following values: For the time domain, we obtained ${\cal PI}=0.53\pm 0.06$ for time series A and $0.52\pm0.05$ for time series B from \sgrb, as well as ${\cal PI}=0.47\pm0.11$ for time series C from \sgra.
In the energy domain, the values were ${\cal PI}=0.68\pm0.09$ for time series A and $0.78\pm0.16$ for time series B from \sgrb, and ${\cal PI}=0.82\pm0.16$ for time series C from \sgra.
We also conducted experiments with $m$ values of $3$--$5$, confirming marginal deviations 
in the Pincus Index values (deviations $\lesssim 7$\%). 
These PI values in magnetar bursts are significantly different in time and energy domains, with the time domain less random than the energy domain.

\subsection{Largest Lyapunov Exponent} 
\label{subsec:LLE}

While there is no universally accepted single definition of chaos, a common measure to quantify sensitivity to initial conditions (i.e., stability) in nonlinear systems is the largest Lyapunov exponent (LLE). The LLE represents the average exponential rate at which even tiny perturbations in a system's state grow or diminish over time. A negative LLE suggests stable dynamics with decreasing uncertainty, while a positive value indicates unstable behavior, and is
widely used as an effective definition of chaos.

%Let us take two points in the time series, $u_i$ and $u_j$, with values that are quite close. This implies that the system reached nearly the same state during the $i$-th and $j$-th iterations. Now, consider the sequences $u_i$, $u_{i+1}$, $u_{i+2}\ldots$ and $u_j$, $u_{j+1}$, $u_{j+2}\ldots$  To quantify how these two sequences diverge from each other, one can calculate the distance between the two sequences after $k$ steps: $d(k)$. In the case of a chaotic system, $d(k)$ initially grows exponentially as $k$ increases. One can plot $\ln d(k)$ against $k$ and then apply a linear fit to the data. The slope of this linear fit provides an estimate of the Lyapunov exponent. 

We use NOnLinear measures for Dynamical Systems ({\fontfamily{qcr}\selectfont nolds}), a Python-based module which provides the algorithm of \citet{Eckmann1986} ({\fontfamily{qcr}\selectfont nolds.lyap\_e}) to estimate the LLE. Our calculations are carried out with the default parameter settings, ensuring consistency with the approach behind Z23 (Y-.K. Zhang in private communication). We calculated two LLE values for $\Delta T_i$ and $\Delta E_i$ separately, following a similar approach to the PI. However, it is important to note that LLE is not a distribution; it represents the maximum value within the vector for a given dataset, making it challenging to define its uncertainty. Therefore, we consider LLE as a rough indicator of the degree of randomness.
As a result, in the time domain, we obtained ${\cal LLE}=0.059$ for time series A and $0.046$ for time series B from \sgrb, while time series C from \sgra exhibited ${\cal LLE}=0.11$. In the energy domain, the values were ${\cal LLE}=0.086$ for time series A and $0.16$ for time series B from \sgrb, with time series C from \sgra displaying ${\cal LLE}=0.16$.
These positive LLE values indicate the presence of significant chaos in magnetar bursts.

\begin{figure*}
    \centering
    \includegraphics[scale=0.52]
%    [width=.48\textwidth]
    {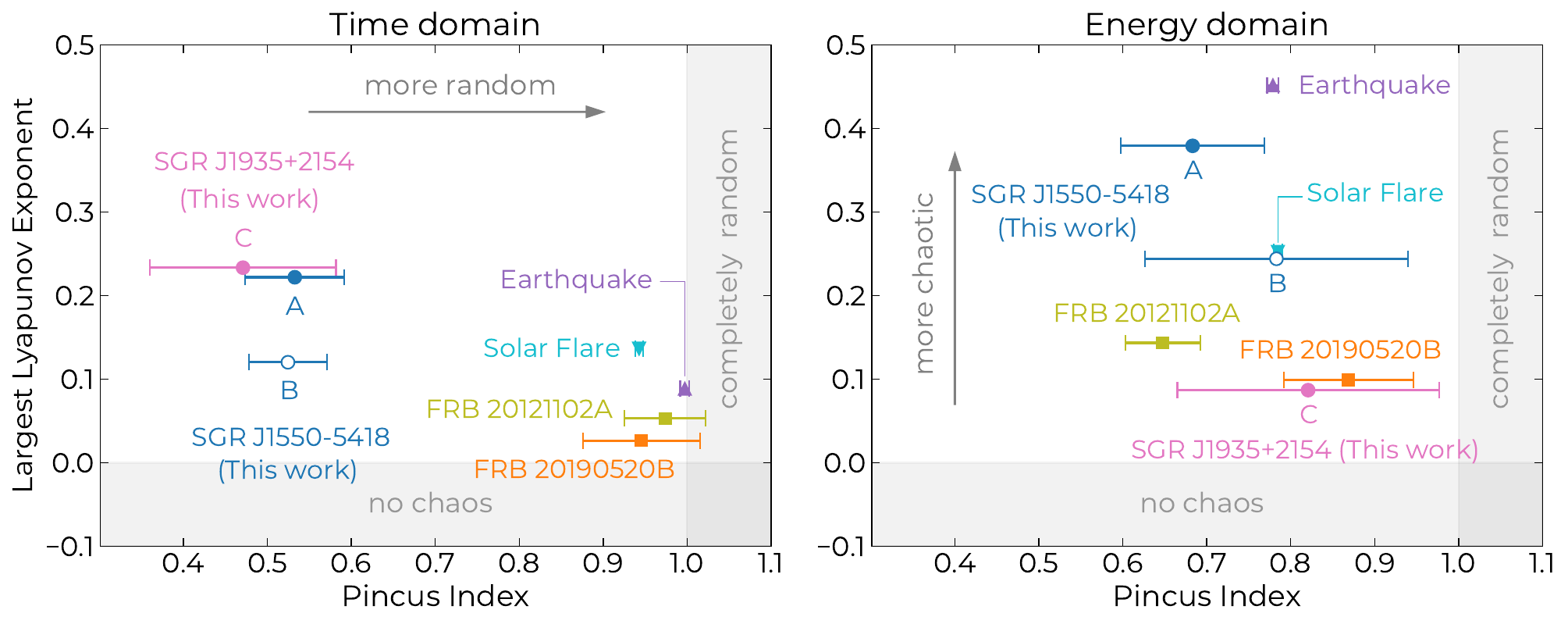}
    \caption{Chaos-randomness plane: Pincus Index vs. Largest Lyapunov Exponent for various phenomena in the time (left) and energy (right) domains. The magnetar bursts newly analyzed in our study are represented by circle symbols.
    The data points for the earthquakes, solar flares, and two repeating FRB sources are obtained by re-analyzing the same datasets adopted from \citet{Zhang2023}. The regions shaded in grey represent extreme phase spaces where phenomena neither exhibit chaos nor show complete randomness. }
    \label{fig:comparison}
\end{figure*}

\section{Discussion \& Summary} 
\label{sec:discussion}

Figure \ref{fig:comparison} illustrates the chaos-randomness phase space, drawing a comparison between magnetar bursts and other phenomena. In this work, only sudden transient phenomena (earthquake, solar flares, and repeating FRBs\footnote{Repeating FRB 20121102A compared here is known to have a quasi-periodic activity window spanning 160 days \citep{Cruces2021}. However, our specific focus here is the burst behavior occurring on a significantly shorter timescale of 10 hours (Z23).}) analyzed in Z23 are compared with magnetar bursts, as their behaviors in time and energy domains are not trivial, making it meaningful to independently compare.  Notably, a clear separation emerges in the time domain, where magnetar bursts are noticeably less random than other phenomena. Importantly, this result remains robust against potential uncertainties in the LLE values, as the difference in PI is statistically significant. 
Conversely, in the energy space, we observe a more uniform distribution of PI values across various phenomena. However, the LLE values in both domains provide limited insights into the differentiation between phenomena, primarily suggesting that magnetar bursts could generally display a wider range of chaos compared to FRBs.
Henceforth, we focus our discussion on the results within the time domain.

Surprisingly, our analysis reveals that both \sgra and \sgrb do not display significant proximity to FRBs in the time domain.  This observation leads us to ponder intriguing questions: Could FRBs be triggered by phenomena beyond the realm of typical magnetar bursts? Notably, the magnetar bursts from \sgrb associated with the Galactic FRB-like event (FRB 200428)  on April 28, 2020 \citep{Stare2,Chime}, exhibited unique spectral characteristics. While typical magnetar bursts (analyzed in this study) usually feature quasi-thermal X-ray spectra peaking around 1--10 keV, the distinctive bursts linked to FRB 200428 displayed exceptionally high peak temperatures at 80 keV \citep{integral_paper,hxmt_paper,konus_paper,agile_paper}. Moreover, the FRB-like event was detected in just one instance among numerous typical magnetar bursts from the same source \citep{Lin2020_nat}.
Therefore, the deviation of typical magnetar bursts from FRBs in the PI-LLE plane may suggest that FRBs (at least two repeating sources) may not be linked to typical bursts but rather to special magnetar bursts, or there could be additional FRB triggering mechanisms that may operate simultaneously with magnetar burst triggers.
In the context of the magnetar-FRB scenario, the generation \citep{Murase2016,Metzger2019,Lu2020} and escape \citep{Ioka2020,Katz2020,Beloborodov2022,Yamasaki2022,Wada2023} of FRBs could be influenced by the energy of associated magnetar flares. 
To address this, exploring how the proximity of magnetar bursts and FRBs shifts based on magnetar burst energy could potentially offer valuable insights into the emission mechanism. However, due to the limited statistics, we defer this for future follow-up studies.

We acknowledge a couple of potential limitations in our discussion. Firstly, the relatively small sample size for magnetar sources/datasets could impact the robustness of our discussion. To establish whether a distinct distribution in the chaos-randomness phase space exists for these phenomena, a more extensive dataset across magnetar sources is crucial.
Secondly, it is important to note that our analysis, while considering randomness and chaos for both time series ($\Delta T_i$) and energy series ($\Delta E_i$), effectively treats them as independent 1D sequences, rather than being analyzed in a true 2D manner where time series and energy series are simultaneously handled. In this regard, our approach based on Z23 differs from the correlation function analysis in a 2D space of time and energy recently conducted by \citet{Totani2023} 
(note that magnetar bursts were not analyzed in their analysis). Remarkably, \citet{Totani2023} found that repeating FRBs share more similarities with earthquakes than solar flares, a finding that contrasts with the results of the dynamical stability analysis by Z23
(see also Figure \ref{fig:comparison_}). Our examination reveals that when considered in the time and energy domains separately, earthquakes are positioned closer to FRBs than solar flares (see the left panel of Figure \ref{fig:comparison}), which qualitatively aligns with the findings of \citet{Totani2023}. However, in the energy domain (see the right panel of Figure \ref{fig:comparison}), the opposite conclusion emerges (inconsistent with \citealt{Totani2023}).
Nonetheless, a direct comparison between our results to their study is challenging due to the employment of distinct methodologies. For example, \citet{Totani2023} employed simulated data assuming a Poisson process, even though not all processes in comparison may necessarily follow a Poisson process (Z23). Thirdly, our approach based on Z23 is relatively novel in the context of astrophysical transients, and further studies involving datasets from various phenomena and sources are warranted to understand and characterize them. Finally, conducting numerical simulations of astrophysical phenomena, specifically magnetar bursts and FRBs, with given levels of randomness and chaos, and then analyzing how these simulation results manifest on the chaos-randomness plane (even further exploration, including the consideration of observational biases), could have substantial implications. These aspects are beyond the scope of our investigation here and could be examined in future studies. 

%Lastly, due to the potential bias arising from the finite data length ($N$), utilizing the maximum value of ApEn with fixed $N$ has been suggested as a more robust complexity estimator for the system. However, it is important to note that $N$ is not held constant across different phenomena in both \citet{Zhang2023} and this study. 

In summary, we explore the dynamical stability of magnetar bursts within the chaos-randomness phase space for the first time. We incorporate burst energy time series data from two active magnetar sources \sgra and \sgrb.  We find distinctive patterns of magnetar bursts compared to various astrophysical phenomena, including enigmatic FRBs. In the time domain, magnetar bursts exhibit a significantly low degree of randomness, whereas in the energy domain, we do not find a significant difference between magnetar bursts and other phenomena. 
Surprisingly, neither \sgra nor \sgrb bursts show significant proximity to repeating FRBs. The deviation of typical magnetar bursts from FRBs in the PI-LLE plane suggests that FRBs are not associated with typical magnetar bursts but may be linked to special magnetar bursts, such as the spectrally peculiar magnetar X-ray bursts observed simultaneously with Galactic FRB 200428.

%While the dynamical similarity between magnetar bursts and FRBs was revealed in this work, it is also possible that not all magnetar bursts are associated with FRBs. For instance, the FRB-like event from \sgrb was detected in just one instance among numerous X-ray bursts \citep{Lin2020_nat}. If this is the case, there still remains a possibility that typical magnetar bursts may not be related to FRB production. 

\section*{Acknowledgements}
We would like to express our gratitude to Di Li and Yongkun Zhang for their valuable discussions and for generously sharing their custom code and comparison data for dynamical stability analysis. 
We also thank Yuki Kaneko for providing the list of SGR J1550$-$5418 bursts and Tomonori Totani for the discussion. We thank the referee for providing a valuable suggestion to consider uncertainties in the result, which greatly helped improve the quality of the manuscript. TH acknowledges support from the National Science and Technology Council of Taiwan through grants 110-2112-M-005-013-MY3, 110-2112-M-007-034-, and 112-2123-M-001-004-.

%%%%%%%%%%%%%%%%%%%%%%%%%%%%%%%%%%%%%%%%%%%%%%%%%%

\section*{Data Availability}
The data of magnetar bursts including their arrival time and fluence are available upon reasonable request to the corresponding author. 



\bibliographystyle{mnras}
\input{sgr_randomness.bbl}

%%%%%%%%%%%%%%%%%%%%%%%%%%%%%%%%%%%%%%%%%%%%%%%%%%

%%%%%%%%%%%%%%%%% APPENDICES %%%%%%%%%%%%%%%%%%%%%

\appendix

\section{Remarks on time-energy averaged results}
\label{sec:caution}

\begin{figure}
    \centering
    \includegraphics[width=.49\textwidth]
    {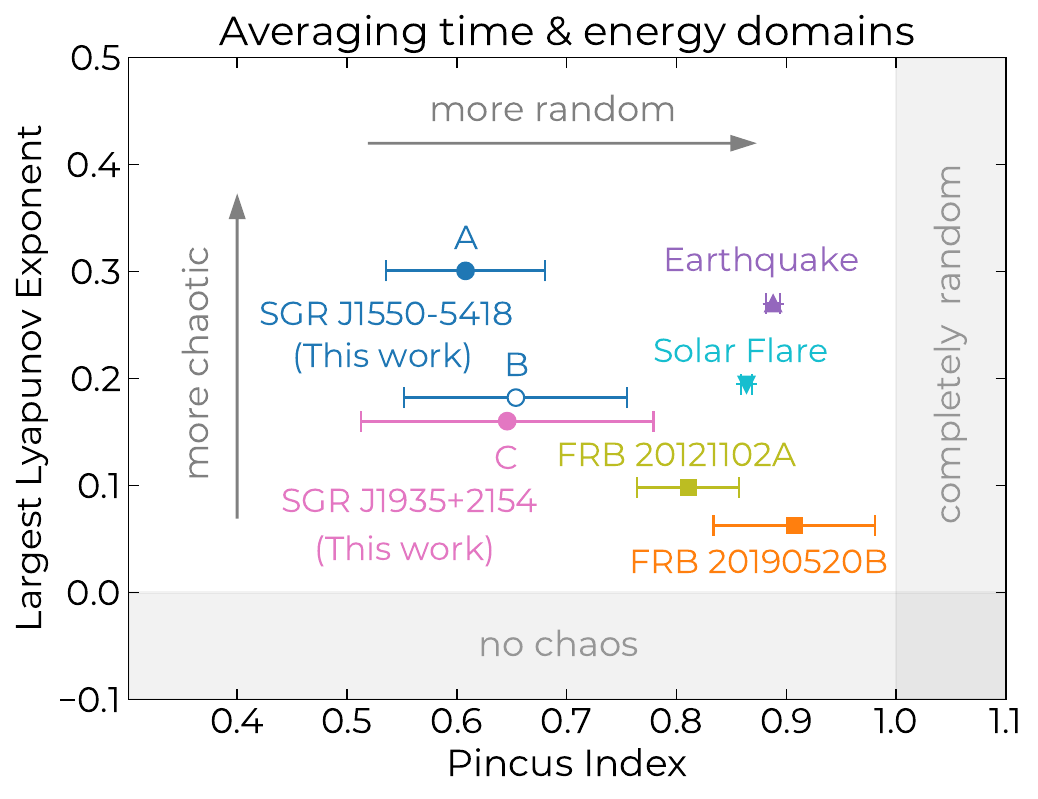}
    \caption{
Chaos-randomness plane: An averaged representation of the data shown in Figure \ref{fig:comparison}, combining results from both time and energy domains, following the original methodology described in Z23 \citep{Zhang2023}.}
    \label{fig:comparison_}
\end{figure}

Figure \ref{fig:comparison_} illustrates the ``averaged'' outcome of the PI-LLE plane, following the methodology initially proposed by Z23, which combines the results of the time series and the energy series.  Clearly, the positions of various phenomena in the PI-LLE phase space can vary significantly from their original positions in the time and energy domains shown in Figure \ref{fig:comparison}.
For instance, on the time-energy averaged plane, solar flares are positioned closer to FRBs than earthquakes. However, in the time domain alone, the situation is reversed (see the left panel of Figure \ref{fig:comparison}). Likewise, the Pincus index values for FRB 20121102A and FRB 20190520B in the left panel of Figure \ref{fig:comparison} are notably higher compared to those in the right panel of Figure \ref{fig:comparison}. This distinction might not be apparent in the time-energy averaged domain illustrated in Figure \ref{fig:comparison_}.
Therefore, we advise general caution when interpreting these results, as a simple averaging approach could potentially lead to misleading conclusions. Additionally, it is crucial to consider the uncertainty associated with the PI to differentiate the degree of randomness, especially as this uncertainty can be significantly large for certain phenomena.

%%%%%%%%%%%%%%%%%%%%%%%%%%%%%%%%%%%%%%%%%%%%%%%%%%

% Don't change these lines
\bsp	% typesetting comment
\label{lastpage}
\end{document}